\documentclass[preprint,showpacs,amsmath,amssymb]{revtex4-1}
\usepackage{graphicx}

\begin{document}
\title{Nonequilibrium transport in molecular junctions with strong electron-phonon interactions}
\author{R. C. Monreal, F. Flores and A. Martin-Rodero}
\affiliation{Departamento de F\'{i}sica Te\'orica de la Materia 
Condensada C05, 
Universidad Aut\'{o}noma de Madrid, Francisco Tom\'as y Valiente,7,
 E-28049 Madrid, Spain}
\begin{abstract}
We present a combined theoretical approach to study the nonequilibrium transport properties of nanoscale systems coupled to metallic electrodes and exhibiting strong electron-phonon interactions.
 We use the Keldysh Green function formalism to generalize  beyond linear theory in the applied voltage an equation of motion method and an
interpolative self-energy approximation  previously developed in equilibrium. 
We analyze the specific characteristics of inelastic transport appearing 
in the intensity versus voltage 
curves and in the conductance, providing qualitative criteria for the sign of the
step-like features in the conductance. 
Excellent overall agreement between both approaches is found for a 
wide range of parameters.
\end{abstract}
\pacs{73.63.-b, 71.38.-k, 73.63.Kv}
\maketitle

\section{Introduction}

Advances in the field of molecular electronics and nano-objects \cite{Reed}
have motivated an increasing
interest in electron-phonon interaction \cite{review}.
Experiments  give evidence that
electron-vibrational coupling within the molecule play an
important role in its charge transport properties. This was
first found by Park et al. on $C_{60}$ \cite{Park}. Also,
the excitation spectra show features that could be
ascribed to sidebands formed by the presence of strong
electron-phonon interactions \cite{Park1,Zhit}.

From the theoretical point of view, the problem of the
interaction of a localized level with a field of bosons can be
traced back to the small polaron model of Holstein \cite{Hols} .
Today the so-called
Anderson-Holstein Hamiltonian is the simplest and more commonly used Hamiltonian
to study the electronic transport through molecular systems.
This Hamiltonian has not an  exact solution
except for a few special cases in equilibrium.
Therefore it is desirable to develop different theoretical approaches which would allow to
calculate and predict robust behaviors for physical magnitudes directly comparable
to out of equilibrium experiments.
With this aim
we use a Keldysh Green function formalism \cite{Keldysh} to generalize two theoretical
approaches previously developed by us in equilibrium,
the equation of motion (EOM) method \cite{EOM}
and the interpolative self-energy approximation (ISA) \cite{ISA}, 
to deal with situations in which many phonons can be
absorbed/emitted by the molecular system  when a bias voltage is applied between the electrodes.
This is clearly a nonequilibrium situation which cannot be described by extensions
of equilibrium theories to small voltages, if the voltage
exceeds the phonon frequency.

Previously, the problem of the electronic transport through molecular junctions
or quantum dots 
has been approached in different ways depending on the different
regimes determined by the parameters: the temperature T, the coupling of
the localized level to the leads characterized by the level width
$\Gamma$, the coupling of the localized
level to phonons $\lambda$, and the phonon frequency $\omega_0$.
The semi-classical regime, defined by $T>>\Gamma$, can be described from
a master equations point of view \cite{Braig,Mitra}. In the
quantum regime $T<<\Gamma$, the ratio $\lambda/\Gamma$
distinguishes between the weak and strong coupling regimes.
The weak coupling regime, $\lambda/\Gamma<<1$, can be approached by a
variety of methods with the common characteristic of being
perturbative in $\lambda/\Gamma$ such as the Born approximation or
the self-consistent Born approximation \cite{Nico,GRN,Yam,Vega,Galp},
perturbative renormalization theory \cite{Paaske} or diagrammatic
techniques \cite{Mitra,Anda,Konig}.
In this respect we should mention that the two approaches introduced in the present
work recover this limit. 
The quantum, strong coupling
 regime, $\lambda/\Gamma>>1$, for which perturbation theory
breaks down is much more difficult to analyze and the decoupling
of electronic and vibronic degrees
of freedom has been a usual approximation
\cite{Galp,Korea,Lun,China}. This work concentrates in this limit approaching
the problem from two very different
 starting points.

A remarkable experimental result is the observed step-like feature in the differential conductance 
at bias voltages equal to the phonon energy that can be either upwards or downwards
\cite{Park,Park1,Zhit,Vega,Park2,Many,Smit}. 
This fact has attracted a considerable theoretical interest \cite{review,GRN} lately.
In the limit 
$\omega_0 << \Gamma$, a symmetric contact and small $\lambda$ the behavior turns out to
depend only on the transmission $\tau$ of the junction with the step upwards (downwards) for
$\tau<1/2$ ($\tau>1/2$) \cite{Vega,Danish}. This result seems to offer a rough
rule of thumb for predicting the observed step sign. Outside this limiting situation this
feature on the conductance will depend in a more complicated 
form on the system parameters \cite{Egger,Imry}. The same issue will be addressed in this work
in the strong coupling regime.

In order to introduce the method we will consider in this paper the spinless version of
the Anderson-Holstein model \cite{Glaz,HN1,HN2}. 
In section I we introduce the nonequilibrium Green functions formalism used for the 
calculation of the transport properties of the system.
In sections II and III we present the out of equilibrium  extensions of the
EOM method and the ISA respectively.
Section IV is devoted to the analysis of the intensity versus voltage curves and the conductance
as obtained by both approximations. The remarkable overall agreement found between such different
theoretical approaches in the out of equilibrium situation for a wide range of parameters,
gives confidence in our results.
We find the I-V curves to increase stepwise when a new inelastic channel emitting n phonons opens.
The conductance reveals more interesting features of the emission processes.
While its main peak, obtained at low voltages, is almost identical to the main resonance appearing in the equilibrium density of states, 
the phonon side-bands show specific behavior associated
to inelastic transport which therefore cannot be obtained by any extension of equilibrium
calculations to finite voltages. The origin of such features is analyzed.
Finally, our conclusions are presented in section V.
Atomic units $e=\hbar=m=1 $ are used  throughout this work except otherwise stated.

\section{General nonequilibrium formalism}

We consider the spinless Anderson-Holstein Hamiltonian describing a single non-degenerate electronic level, $\epsilon_0$, coupled linearly to a local phonon mode of frequency $\omega_0$ and to electronic reservoirs, 
 
\begin{equation}
\hat{H} =
\epsilon_0\hat{c}_{0}^{\dagger}\hat{c}_{0}+\sum_{k,\nu}\left(V_{k,\nu}\hat{c}_{k,\nu}^{\dagger}\hat{c}_{0}+c.c.\right) +\sum_{k,\nu}(\epsilon_{k,\nu}+\mu_{\nu})\hat{c}_{k,\nu}^{\dagger}\hat{c}_{k,\nu}
+\omega_{0}\hat{b}^{\dagger}\hat{b}+\lambda(\hat{b}^{\dagger}+\hat{b})\hat{n}_{0}
\label{Holstein}
\end{equation} 
where $\omega_0$ is the phonon energy, $\lambda$ the electron-phonon coupling constant, $\epsilon_{k,\nu}$ with $\nu=L,R$ denotes the single particle energies of the left and right electrodes, $\mu_{L}-\mu_{R}=eV$ being the applied bias and $V_{k,\nu}$ the coupling between the localized level and the reservoir states.

The electronic transport properties through this system can be conveniently calculated using the nonequilibrium Green function formalism or Keldysh method \cite{Keldysh}. For a stationary situation the retarded $G^{r}$ and the nonequilibrium distribution Green functions $G^{+-}$ and $G^{-+}$ are defined as follows: 
\begin{eqnarray}
G_{ij}^{r}(\omega)&=&-i\int\theta(t-t')<c_{j}^{\dagger}(t)c_{i}(t')+c_{i}(t')c_{j}^{\dagger}(t)>e^{i\omega(t-t')}d(t-t') \nonumber \\
G_{ij}^{+-}(\omega)&=&i\int<c_{j}^{\dagger}(t)c_{i}(t')>e^{i\omega(t-t')}d(t-t') \\
G_{ij}^{-+}(\omega)&=&-i\int<c_{i}(t')c_{j}^{\dagger}(t))>e^{i\omega(t-t')}d(t-t') \nonumber
\label{Keldysh-Green1}
\end{eqnarray}

The frequency dependent Keldysh Green functions can be obtained from the corresponding Dyson equations which in matrix form read:

\begin{eqnarray}
\bf{G^{+-}}&=&\bf{g^{+-}}+\bf{g^{+-}}\Sigma^{a}G^{a}+g^{r}\Sigma^{r}G^{+-}-g^{r}\Sigma^{+-}G^{a} \\ \nonumber
\bf{G^{+-}}&=&\bf{g^{+-}}+G^{+-}\Sigma^{a}g^{+-}+G^{r}\Sigma^{r}g^{+-}-G^{r}\Sigma^{+-}g^{a} \\ 
\bf{G^{+-}}&=&(\bf{I}+\bf{G^{r}}\bf{\Sigma^{r}})g^{+-}(I+\Sigma^{a}G^{a})-G^{r}\Sigma^{+-}G^{a} \nonumber
\label{Keldysh-Green2}
\end{eqnarray}
where $\bf{I}$ is the unit matrix and $\bf{g}$ are the Green functions of the uncoupled system ($V_{k,\nu}=0$) and with  similar equations for the $G^{-+}$ functions. The crucial point within this formalism consists in finding a reasonable approximation for the self-energies $\Sigma^{r}$ and $\Sigma^{+-}$.
The current intensity between the reservoir $\nu$ and the quantum level can be written in terms of the $G^{+-}$ Green functions as:
\begin{equation}
I_{\nu}=\frac{e}{h}\sum_{k}V_{k,\nu}\int d\omega [G_{k\nu,0}^{+-}(\omega)-G_{0,k\nu}^{+-}(\omega)]
\label{current}
\end{equation}
where the subindex $0$ labels the dot level.

Using Eqs.({\ref{Keldysh-Green2}) it is possible to write the current density in terms of the dot level Green functions. In particular, Eqs. (3) lead to \cite{Caroli}:
\begin{equation}
\sum_{k}V_{k,\nu}[G_{k\nu,0}^{+-}-G_{0,k\nu}^{+-}]=\sum_{k}|V_{k,\nu}|^{2}[g_{kk,\nu}^{+-}G_{00}^{-+}-G_{00}^{+-}g_{kk,\nu}^{-+}]
\label{G001}
\end{equation}
where the Green function $G_{00}^{+-}$ is calculated from the corresponding Dyson equation:
\begin{equation}
G_{00}^{+-}=\sum_{k,\nu}|V_{k,\nu}|^{2}|G_{00}^{r}|^{2}g_{kk,\nu}^{+-}-|G_{00}^{r}|^{2}\Sigma_{00}^{+-}
\label{G002}
\end{equation}
with a similar equation for $G_{00}^{-+}$. All the expressions can be simplified by making the usual wide-band approximation \cite{Wingreen}:

\begin{equation}
\sum_{k}|V_{k}^{\nu}|^{2}g_{kk,\nu}^{+-}(\omega)=2i\Gamma_{\nu}f_{\nu}(\omega)
\end{equation}
where $f_{\nu}(\omega)$ are the Fermi distribution functions of the electrodes and $\Gamma_\nu$ are taken as constants.

From Eqs.(\ref{G001}) and (\ref{G002}) the current can be written as a sum of an elastic and an inelastic contribution, $I_{\nu}=I_{\nu}^{(el)}+I_{\nu}^{(in)}$ as:

\begin{eqnarray}
I_{L,R}^{(el)}&=&\frac{4e}{h}\Gamma_{L}\Gamma_{R}\int d\omega |G_{00}^{a}(\omega)|^{2}[f_{L,R}(\omega)-f_{R,L}(\omega)] \nonumber \\
I_{\nu}^{(in)}&=&-\frac{2ie}{h}\Gamma_{\nu} \int d\omega 
|G_{00}^{a}(\omega)|^{2}[\Sigma_{00}^{-+}(\omega)f_{\nu}(\omega)+\Sigma_{00}^{+-}(\omega)(1-f_{\nu}(\omega))]
\label{current2}
\end{eqnarray}

Due to current conservation, $I_{L}=-I_{R}$,
an equivalent expression can be obtained by means of the identity $I=(\Gamma_{R}I_{L}-\Gamma_{L}I_{R})/\Gamma$ with $\Gamma=\Gamma_{L}+\Gamma_{R}$ leading from Eqs. (\ref{current2}) to the well known expression \cite{Wingreen}:
\begin{equation}
I=\frac{e}{h}\frac{4\Gamma_{L}\Gamma_{R}}{\Gamma}\int d\omega ImG_{00}^{a}(\omega)[f_{L}(\omega)-f_{R}(\omega)] 
\label{current-Wingreen}
\end{equation}

From Eq.(\ref{current-Wingreen}), the differential conductance is obtained as
$G=dI/dV$.
In the linear response regime $V\rightarrow 0$, $G_{00}^{a}(\omega)$ can be evaluated in
equilibrium, $G_{00}^{a}(\omega)\simeq G_{00}^{a,eq}(\omega)$, 
and the conductance can be expressed in terms of
$ImG_{00}^{a,eq}(\omega=\mu_L)$ and $ImG_{00}^{a,eq}(\omega=\mu_R)$. However, this 
is not in general a good approximation for $V\geq \omega_0$ 
and a full calculation of $G_{00}^{a}(\omega)$ has to be performed.

On the other hand, the level occupation can be obtained from the non equilibrium spectral density functions as:
\begin{eqnarray}
\frac{1}{2\pi i}\int d\omega G_{00}^{+-}(\omega)&=&<n> \nonumber \\
\frac{1}{2\pi i}\int d\omega F_{00}(\omega)&=&2<n>-1
\label{occupation}
\end{eqnarray}
where $F_{00}=G_{00}^{+-}+G_{00}^{-+}$.

\section{Equation of Motion method for the Anderson-Holstein Hamiltonian out of equilibrium}
In the quantum strong coupling regime we are interested in, it is 
convenient to apply to Hamiltonian Eq.(1) a standard 
canonical transformation  
$\tilde{H}=\hat{S}\hat{H}\hat{S}^{-1}$  with $\hat{S}$ given by
\cite{Holstein-canonical,Mahan}
\begin{equation}
\hat{S}=exp[\frac{\lambda}{\omega_0}({b}^{\dagger}-{b})\hat{n}_{0}]
\label{eq-S}
\end{equation}

which transforms electronic and bosonic operators as

\begin{eqnarray}
\tilde{c}_0 & = & \hat{c}_0
exp[-\frac{\lambda}{\omega_0}({b}^{\dagger}-{b})]
\nonumber \\
\tilde{c}_{k,\nu} & = & \hat{c}_{k, \nu}
\nonumber \\
\tilde{b} & = & {b}-\frac{\lambda}{\omega_0}
\hat{n}_{0}
\nonumber \\
\label{op-tilde}
\end{eqnarray}
Note that Eqs.(\ref{op-tilde}) imply that the number operators for
electrons in the level and in the leads remain unchanged.
Then, the transformed Hamiltonian reads:

\begin{equation}
\tilde{H}= \tilde\epsilon_0 \hat{n}_{0} +
\sum_{k,\nu}\epsilon_{k, \nu}\hat{n}_{k,\nu}+
\sum_{k,\nu} V_{k, \nu} (
\hat{c}_{k,\nu }^{\dagger}\tilde{c}_{0} +
\tilde{c}_{0}^{\dagger}\hat{c}_{k,\nu} )+
\omega_0 {b}^{\dagger}{b}
\label{H-tilde}
\end{equation}
with $\tilde\epsilon_0=\epsilon_0-\lambda^2/\omega_0$
representing the renormalization of the energy level due to
its coupling with the local phonon.

The nonequilibrium Green's functions will also be written in terms of
the tilde-operators.
In the EOM procedure, we will obtain Green's functions for other operators
$\hat{O}(t)$ different from $\tilde{c}_{0}(t)$ 
at time $t$, which are defined in a way similar to Eqs.(\ref{Keldysh-Green1})
but with a more convenient notation. Also, instead of the functions $G^{+-}$ and $G^{-+}$ it is more convenient to use here their sum $F$. Then we write

\begin{eqnarray}
G^{a}(\hat{O};t,t')&=&i\theta(t'-t)
<\hat{O}(t)\tilde{c}_{0}^{\dagger}(t')+\tilde{c}_{0}^{\dagger}(t')
\hat{O}(t)>_{\tilde{H}} \nonumber \\
F(\hat{O};t,t')&=&i
<\tilde{c}_{0}^{\dagger}(t')\hat{O}(t)-
\hat{O}(t)\tilde{c}_{0}^{\dagger}(t')>_{\tilde{H}}
\label{G-F}
\end{eqnarray}

From now on, the symbol $<....>_{\tilde{H}}$  means that the average should be
taken with respect to the transformed Hamiltonian $\tilde{H}$.

The EOM method for solving the Anderson-Holstein Hamiltonian  was  
already introduced in \cite{EOM} . Briefly, starting with 
$G^{a}(\tilde{c}_0; t,t')$ from Eq.(\ref{G-F}) and applying the equation of motion, a 
hierarchy of new Green's functions $G^{a}(b^{\dagger i}\tilde{c}_{0}b^{j};t,t')$ 
is generated. To obtain a closed system, at a given step of the procedure 
we contract pairs of operators
$\hat{c}_{k, \nu}$ and $\hat{c}_{k', \nu}^{\dagger}$ where possible  as

\begin{equation}
\hat{c}_{k', \nu'}^{\dagger}\hat{c}_{k, \nu}\simeq \delta_{k,k'}
\delta_{\nu, \nu'}<n_{k, \nu}>
\label{contract}
\end{equation}
In this equation 
 $<n_{k, \nu}>$ is the Fermi-Dirac distribution function of the 
$\nu$-electrode.
Due to the fact that the non equilibrium problem we are interested in 
is much more involved than the equilibrium one addressed in \cite{EOM},
 we will restrict the method to the order $O(V_{k,\nu}^2)$.
Then, the following system of linear equations has to be solved:

\begin{eqnarray}
(\omega_{ij}-\tilde\epsilon_{0}-\Gamma(\omega_{ij}))
G^{a}(b^{\dagger i} \tilde{c}_{0} b^{j};\omega) = 
<\tilde{c}_{0}^{\dagger}b^{\dagger i}\tilde{c}_{0}b^{j}
+b^{\dagger i}\tilde{c}_{0}b^{j}\tilde{c}_{0}^{\dagger}>_{\tilde{H}} & +  &  
\nonumber \\
\sum_{l=0}^{i}
\left( \begin{array}{l} i\\l \end{array} \right)
\left(-\frac{\lambda}{\omega_0} \right)^{i-l}
\sum_{k, \nu} V_{k, \nu}
\frac{<b^{\dagger l}\tilde{c}_{0}^{\dagger}\hat{c}_{k, \nu}(b+
\frac{\lambda}{\omega_0})^{j}>_{\tilde{H}}}{\omega_{lj}-\epsilon_{k, \nu}-i\eta}& - & 
\nonumber \\
\sum_{l=0}^{j}
\left( \begin{array}{l} j\\l \end{array} \right)
\left(\frac{\lambda}{\omega_0} \right)^{j-l}
\sum_{k, \nu} V_{k, \nu}
\frac{<(b^{\dagger}-\frac{\lambda}{\omega_0})^{i}\tilde{c}_{0}^{\dagger}\hat{c}_{k, \nu}
b^{j}>_{\tilde{H}}}{\omega_{il}-\epsilon_{k, \nu}-i\eta}& +& 
\nonumber \\
\lambda G^{a}(b^{\dagger i} \tilde{c}_{0}b^{j+1};\omega)+
\lambda G^{a}(b^{\dagger i+1} \tilde{c}_{0}b^{j};\omega)& +& 
\nonumber \\
\sum_{l=0}^{j-1}\left(-\frac{\lambda}{\omega_0} \right)^{j-l}
G^{a}(b^{\dagger i} \tilde{c}_{0} b^{l};\omega)
\sum_{m=l}^{j}(-1)^{j-m}
\left( \begin{array}{l} j\\m \end{array} \right)
\left( \begin{array}{l} m\\l \end{array} \right)
\Gamma^{(h)}(\omega_{im})&+ & 
\nonumber \\
\sum_{l=0}^{i-1}\left(\frac{\lambda}{\omega_0} \right)^{i-l}
G^{a}(b^{\dagger l} \tilde{c}_{0} b^{j};\omega)
\sum_{m=l}^{i}(-1)^{i-m}
\left( \begin{array}{l} i\\m \end{array} \right)
\left( \begin{array}{l} m\\l \end{array} \right)
\Gamma^{(e)}(\omega_{mj})
\label{eomGa}
\end{eqnarray}
with
 $\left( \begin{array}{l} n\\l \end{array} \right)=\frac{n!}{l!(n-l)!}$. 
We have defined $\omega_{ij}=\omega+(i-j)\omega_0$ and the  advanced self-energies 

\begin{eqnarray}
\Gamma(\omega_{ij}) & = &\sum_{k, \nu}
\frac{V_{k, \nu}^{2}}{\omega_{ij}-\epsilon_{k, \nu}-i\eta}
\nonumber \\
\Gamma^{(e)}(\omega_{ij}) & = & \sum_{k, \nu}  V_{k, \nu}^{2}
\frac {<n_{k, \nu}>}{\omega_{ij}-\epsilon_{k, \nu}-i\eta}
\nonumber \\
\Gamma^{(h)}(\omega_{ij}) & = & \sum_{k, \nu}  V_{k, \nu}^{2}
\frac {1-<n_{k, \nu}>}{\omega_{ij}-\epsilon_{k, \nu}-i\eta}
\label{Gammas}
\end{eqnarray}
$\eta$ being an infinitesimal. In the wide-band limit to be used in this work 
$\Gamma(\omega_{ij})=i(\Gamma_{L}+\Gamma_{R})$. 

We should point out that this procedure does not decouple electrons and phonons as it
has been frequent in the literature. Rather, quantum coherence is preserved in all of
the Green functions 
$G^{a}(b^{\dagger i} \tilde{c}_{0} b^{j};\omega)$ 
which involve emission of $i$ and absorption of $j$ phonons.
On the other hand, since we have decoupled the Green functions involving the localized level 
and the electrodes to the order $O(V_{k,\nu}^2)$, the procedure is somehow perturbative in 
$V_{k, \nu}$. However it becomes exact not only in the limit $V_{k, \nu} \rightarrow 0$ 
but for 
$\lambda\rightarrow 0$ and finite $V_{k, \nu} $ as well.

In Reference \cite{EOM} we argued that all the expectation values of the type
$<b^{\dagger n}\tilde{c}_{0}^{\dagger} \hat{c}_{k,\nu}b^{m}>_{\tilde{H}}$
appearing in Eq.(\ref{eomGa}) 
could be neglected. This is not in general the case when an electric 
current circulates
through the localized level because these expectation values
just describe the transit of an electron from the electrode to the level
with absorption of $m$ and emission of $n$ phonons, which is the process we are analyzing.
Therefore, they have to be calculated consistently with the appropriate non equilibrium
 Green's
functions F's as we will explain below. With respect to 
$<b^{\dagger n}\tilde{c}_{0}^{\dagger} \tilde{c}_{0}b^{m}>_{\tilde{H}}$,
these expectation values describe fluctuations in level occupancy when phonons
are absorbed and emitted and should also be calculated consistently with the
appropriate F's functions. However, we have checked that the approximation

\begin{equation}
<b^{\dagger n}
\tilde{c}_{0}^{\dagger}\tilde{c}_{0}b^{m}>_{\tilde{H}}
\cong \delta_{m0}\delta_{n0}
<\tilde{c}_{0}^{\dagger}\tilde{c}_{0}>
\label{ve}
\end{equation}
is still a good approximation out of equilibrium at zero temperature. 

The calculation of the F's Green's functions follows the same lines
even though it is more involved. Starting from 
$F(\tilde{c}_{0};t,t')$ and applying the equation of motion, we obtain new
Green's functions, which are calculated from their equations of motion.
A typical equation being

\begin{eqnarray}
\frac{dF(b^{\dagger i}\tilde{c}_{0}b^{j};t,t')}{dt}=
-i(\tilde\epsilon_{0}-(i-j)\omega_{0})F(b^{\dagger i}\tilde{c}_{0}b^{j};t,t')& -& 
\nonumber \\
i\lambda F(b^{\dagger (i+1)}\tilde{c}_{0} b^{j};t,t')-
i\lambda F(b^{\dagger i}\tilde{c}_{0}b^{j+1};t,t') &+ &
\nonumber \\
i \sum_{k, \nu} V_{k, \nu}
[F((b^{\dagger}-\frac{\lambda}{\omega_0})^{i}
\tilde{c}_{0}^{\dagger}\tilde{c}_{0} \hat{c}_{k, \nu} b^{j};t,t')+
F(b^{\dagger i} \tilde{c}_{0}\tilde{c}_{0}^{\dagger}
\hat{c}_{k, \nu}
(b+\frac{\lambda}{\omega_0})^{j};t,t')]
\label{eom1}
\end{eqnarray}
In the next step, the F's functions appearing in the forth term of Eq.(\ref{eom1})
are calculated from their EOM and approximated by the contraction of operators
indicated in 
Eq.(\ref{contract}), yielding:

\begin{eqnarray}
 \frac{d}{dt}
F(b^{\dagger n}\tilde{c}_{0}^{\dagger}\tilde{c}_{0} \hat{c}_{k, \nu} b^{m};t,t')
=-i(\epsilon_{k, \nu}-(n-m)\omega_0)
F(b^{\dagger n}\tilde{c}_{0}^{\dagger}\tilde{c}_{0} \hat{c}_{k, \nu} b^{m};t,t') &+ &
\nonumber \\
i V_{k, \nu} <n_{k, \nu}>
F((b^{\dagger}+\frac{\lambda}{\omega_0})^{n}\tilde{c}_{0} b^{m};t,t')
\label{eom2}
\end{eqnarray}
and

\begin{eqnarray}
 \frac{d}{dt}F(b^{\dagger n}
\tilde{c}_{0}\tilde{c}_{0}^{\dagger} \hat{c}_{k, \nu} b^{m};t,t')=
-i(\epsilon_{k, \nu}-(n-m)\omega_0)
F(b^{\dagger n}\tilde{c}_{0}\tilde{c}_{0}^{\dagger} \hat{c}_{k, \nu} b^{m};t,t')& +& 
\nonumber \\
i V_{k, \nu} <1-n_{k, \nu}>
F(b^{\dagger n} \tilde{c}_{0}(b-\frac{\lambda}{\omega_0})^{m};t,t')
\label{eom3}
\end{eqnarray}

Eqs.(\ref{eom2}) and (\ref{eom3}) are now integrated in time from an initial time
$t=t_0$ where the system starts to evolve, with the initial conditions

\begin{equation}
F(b^{\dagger n}\tilde{c}_{0}^{\dagger}\tilde{c}_{0} \hat{c}_{k, \nu} b^{m};t_0,t')=
<2n_{k, \nu}-1>
G^{a}(b^{\dagger n}\tilde{c}_{0}^{\dagger}\tilde{c}_{0} \hat{c}_{k, \nu} b^{m};t_0,t')
\end{equation}
and

\begin{equation}
F(b^{\dagger n}\tilde{c}_{0}\tilde{c}_{0}^{\dagger} \hat{c}_{k, \nu} b^{m};t_0,t')=
<2n_{k, \nu}-1>
G^{a}(b^{\dagger n}\tilde{c}_{0}\tilde{c}_{0}^{\dagger} \hat{c}_{k, \nu} b^{m};t_0,t')
\end{equation}

These equations come from the general definitions of Eq.(\ref{G-F})
by taking into account that, initially, 
 the localized level and the leads were non-interacting independent systems. Since the EOM for the advanced Green's functions were previously derived, they are integrated backwards in time $t$, from its final value $t'$ to its initial
value $t_0$. Then we obtain

\begin{eqnarray}
F(b^{\dagger n}\tilde{c}_{0}^{\dagger}\tilde{c}_{0} \hat{c}_{k, \nu} b^{m};t,t')=
-i<2n_{k, \nu}-1><b^{\dagger n}\tilde{c}_{0}^{\dagger}\hat{c}_{k, \nu}
(b+\frac{\lambda}{\omega_0})^m>_{\tilde{H}}
e^{-i(\epsilon_{k, \nu}-(n-m)\omega_0)(t-t')} &- &
\nonumber \\
iV_{k, \nu}<n_{k, \nu}><2n_{k, \nu}-1>\int_{t_0}^{t'} d\tau
G^{a}((b^{\dagger}+\frac{\lambda}{\omega_0})^{n}\tilde{c}_{0}b^{m};\tau,t')
e^{-i(\epsilon_{k, \nu}-(n-m)\omega_0)(t-\tau)} &+ &
\nonumber \\
iV_{k, \nu} <n_{k, \nu}>\int_{t_0}^{t} d\tau
F((b^{\dagger}+\frac{\lambda}{\omega_0})^{n}\tilde{c}_{0}b^{m};\tau,t')
e^{-i(\epsilon_{k, \nu}-(n-m)\omega_0)(t-\tau)} 
\label{ve1}
\end{eqnarray}

and
\begin{eqnarray}
F(b^{\dagger n}\tilde{c}_{0}\tilde{c}_{0}^{\dagger} \hat{c}_{k, \nu} b^{m};t,t')=
i<2n_{k, \nu}-1><((b^{\dagger}-\frac{\lambda}{\omega_0})^{n}\tilde{c}_{0}^{\dagger}\hat{c}_{k, \nu}
b^m>_{\tilde{H}}
e^{-i(\epsilon_{k, \nu}-(n-m)\omega_0)(t-t')} &- &
\nonumber \\
iV_{k, \nu}<1-n_{k, \nu}><2n_{k, \nu}-1>\int_{t_0}^{t'} d\tau
G^{a}(b^{\dagger n}\tilde{c}_{0}(b-\frac{\lambda}{\omega_0})^{m};\tau,t')
e^{-i(\epsilon_{k, \nu}-(n-m)\omega_0)(t-\tau)} &+ &
\nonumber \\
iV_{k, \nu}<1-n_{k, \nu}>\int_{t_0}^{t} d\tau
F(b^{\dagger n}\tilde{c}_{0}(b-\frac{\lambda}{\omega_0})^{m};\tau,t')
e^{-i(\epsilon_{k, \nu}-(n-m)\omega_0)(t-\tau)}
\label{ve2}
\end{eqnarray}

When $t_0 \rightarrow -\infty$, the Fourier transform of 
Eqs.(\ref{ve1}) and (\ref{ve2}) can  be readily obtained after
taking into account that the integrals appearing in these equations can be written as the 
convolution product of two functions. Eq.(\ref{eom1}) is also Fourier transformed 
yielding the final expression that allows us to obtain the F's Green's functions from:

\begin{eqnarray}
(\omega_{ij}-\tilde\epsilon_{0}-\Gamma^{*}(\omega_{ij}))
F(b^{\dagger i} \tilde{c}_{0} b^{j};\omega)  = 
\Omega(\omega_{ij})G^{a}(b^{\dagger i} \tilde{c}_{0} b^{j};\omega)& +&
\nonumber \\
i 2\pi\sum_{l=0}^{i}
\left( \begin{array}{l} i\\l \end{array} \right)
\left(-\frac{\lambda}{\omega_0} \right)^{i-l}
\sum_{k, \nu} V_{k, \nu} <2n_{k, \nu}-1>
<b^{\dagger l}\tilde{c}_{0}^{\dagger}\hat{c}_{k, \nu}(b+
\frac{\lambda}{\omega_0})^{j}>_{\tilde{H}}\delta(\omega_{lj}-\epsilon_{k, \nu})& - & 
\nonumber \\
i 2\pi \sum_{l=0}^{j}
\left( \begin{array}{l} j\\l \end{array} \right)
\left(\frac{\lambda}{\omega_0} \right)^{j-l}
\sum_{k, \nu} V_{k, \nu}  <2n_{k, \nu}-1>
<(b^{\dagger}-\frac{\lambda}{\omega_0})^{i}\tilde{c}_{0}^{\dagger}\hat{c}_{k, \nu}
b^{l}>_{\tilde{H}}\delta(\omega_{il}-\epsilon_{k, \nu}) & +& 
\nonumber \\
\lambda F(b^{\dagger i} \tilde{c}_{0}b^{j+1};\omega)+
\lambda F(b^{\dagger (i+1)} \tilde{c}_{0}b^{j};\omega)& +& 
\nonumber \\
\sum_{l=0}^{j-1}\left(-\frac{\lambda}{\omega_0} \right)^{j-l}[
G^{a}(b^{\dagger i} \tilde{c}_{0} b^{l};\omega)
\sum_{m=l}^{j}(-1)^{j-m}
\left( \begin{array}{l} j\\m \end{array} \right)
\left( \begin{array}{l} m\\l \end{array} \right)
\Omega^{(h)}(\omega_{im}) & +&   
\nonumber \\
F(b^{\dagger i} \tilde{c}_{0} b^{l};\omega)
\sum_{m=l}^{j}(-1)^{j-m}
\left( \begin{array}{l} j\\m \end{array} \right)
\left( \begin{array}{l} m\\l \end{array} \right)
\Gamma^{(h) *}(\omega_{im})] & + & 
\nonumber \\
\sum_{l=0}^{i-1}\left(\frac{\lambda}{\omega_0} \right)^{i-l}[
G^{a}(b^{\dagger l} \tilde{c}_{0} b^{j};\omega)
\sum_{m=l}^{i}(-1)^{i-m}
\left( \begin{array}{l} i\\m \end{array} \right)
\left( \begin{array}{l} m\\l \end{array} \right)
\Omega^{(e)}(\omega_{mj}) & +&
\nonumber \\
F(b^{\dagger l} \tilde{c}_{0} b^{j};\omega)
\sum_{m=l}^{i}(-1)^{i-m}
\left( \begin{array}{l} i\\m \end{array} \right)
\left( \begin{array}{l} m\\l \end{array} \right)
\Gamma^{(e) *}(\omega_{mj})]
\label{eomF}
\end{eqnarray}
where we have defined the following self-energies:

\begin{eqnarray}
\Omega(\omega_{ij}) & = & i 2\pi \sum_{k, \nu}
V_{k, \nu}^{2} <2n_{k, \nu}-1> \delta(\omega_{ij}-\epsilon_{k, \nu})
\nonumber \\
\Omega^{(e)}(\omega_{ij}) & = & i 2\pi \sum_{k, \nu}  V_{k, \nu}^{2}
<2n_{k, \nu}-1><n_{k, \nu}> \delta(\omega_{ij}-\epsilon_{k, \nu})
\nonumber \\
\Omega^{(h)}(\omega_{ij}) & = & i 2\pi \sum_{k, \nu}  V_{k, \nu}^{2}
<2n_{k, \nu}-1> <1-n_{k, \nu}> \delta(\omega_{ij}-\epsilon_{k, \nu})
\label{Omegas}
\end{eqnarray}

The linear sets of Eqs. (\ref{eomGa}) and (\ref{eomF}) are coupled trough the different
expectation values appearing in these equations, which have to be calculated self-consistently.
To do so, notice that from the definitions of
$F(b^{\dagger n}\tilde{c}_{0}^{\dagger}\tilde{c}_{0} \hat{c}_{k, \nu} b^{m};t,t')$
and
$F(b^{\dagger n}\tilde{c}_{0} b^{m};t,t')$
and for $t=t' \rightarrow +\infty$ one  has the identities

\begin{equation}
\int_{-\infty}^{+\infty} \frac{d\omega}{2\pi}
F(b^{\dagger n}\tilde{c}_{0}^{\dagger}\tilde{c}_{0} \hat{c}_{k, \nu} b^{m};\omega)  = 
F(b^{\dagger n}\tilde{c}_{0}^{\dagger}\tilde{c}_{0} \hat{c}_{k, \nu} b^{m};t',t')=
i <b^{\dagger n}\tilde{c}_{0}^{\dagger}\hat{c}_{k, \nu}(b+
\frac{\lambda}{\omega_0})^{m}>_{\tilde{H}}  
\label{veck}
\end{equation}
and

\begin{equation}
\int_{-\infty}^{+\infty} \frac{d\omega}{2\pi} F
(b^{\dagger n}\tilde{c}_{0} b^{m};\omega)  =
F(b^{\dagger n}\tilde{c}_{0} b^{m};t',t')=
i <b^{\dagger n}\tilde{c}_{0} \tilde{c}_{0}^{\dagger}(b+
\frac{\lambda}{\omega_0})^{m} + 
(b^{\dagger}-\frac{\lambda}{\omega_0})^{n}\tilde{c}_{0}^{\dagger}\hat{c}_{0}
b^{m}>_{\tilde{H}}
\label{vec0}
\end{equation}
respectively. By making use of these relations we can obtain the required expectation values
from the EOM of the F's Green's functions. 
 Once the system of Eqs.(\ref{eomGa}) and (\ref{eomF}) are solved, the current $I$ is calculated from Eq.(\ref{current-Wingreen}). An important point is related to current
conservation, $I_{L}=-I_{R}$, which is not automatically satisfied for a given approximation
(see \cite{Hershfield,dotne}). 
We have numerically checked that the EOM method fulfills current conservation within the accuracy of the calculation, in the range of parameters investigated in the present work.

\section{Interpolative solution for the Anderson-Holstein Hamiltonian out of equilibrium}

In this section we will introduce an interpolative approach for the calculation of the self-energy out of equilibrium. This approach is a generalization of a previous one developed for an equilibrium situation \cite{ISA}. It has also been successfully applied to a purely electronic problem like a quantum dot out of equilibrium \cite{dotne}. An interpolative approach is possible due to a property of the self-energy which exhibits the same mathematical form when expanded  in the interaction parameter \cite{interpol1,interpol2,interpol3} (which in Hamiltonian (\ref{Holstein}) is the electron-phonon coupling $\lambda$) both in the atomic ($V_{k,\nu}\rightarrow 0$) an in the perturbative ($\lambda\rightarrow0$) limit. 

We briefly summarize the interpolative approach in an equilibrium situation.
In the $V_{k,\nu}\rightarrow0$ limit Eq.(\ref{Holstein}) can be exactly diagonalized by means of a canonical transformation \cite{Holstein-canonical,Mahan} yielding for the level retarded Green function:
\begin{equation}
G_{00}^{(at)}(\omega)=e^{-\frac{\lambda^2}{\omega_{0}^2}} \sum_{m=0}^{\infty}\frac{\lambda^{2m}}{\omega_{0}^{2m}m!}\left(\frac{1-<\hat{n}>}{\omega-\tilde{\epsilon}_{0}-m\omega_{0}+i\eta}+\frac{<\hat{n}>}{\omega-\tilde{\epsilon}_{0}+m\omega_{0}+i\eta}\right)
\label{atomic-G}
\end{equation}
where $\tilde{\epsilon}_0=\epsilon_0-\lambda^2/\omega_0$ and $\langle \hat{n} \rangle$ is the level occupation. From Eq. (\ref{atomic-G}) we can calculate the expression for the level self-energy by means of the corresponding Dyson equation $\Sigma_{00}^{(at)}=\omega-\epsilon_H -G_{00}^{(at)-1}$ where $\epsilon_H=\epsilon_0-2(\lambda^2/\omega_0)\langle n \rangle$ is the energy level corrected by the Hartree contribution. In the limit of small electron-phonon coupling $\lambda/\omega_{0}<<1$ and to order $\lambda^{2}$, the atomic self-energy tends to:
\begin{equation}
\Sigma_{00}^{(at)}(\omega)\approx \lambda^2\left(\frac{1-<\hat{n}>}{\omega-\epsilon_0-\omega_0+i\eta}+\frac{<\hat{n}>}{\omega-\epsilon_0+\omega_0+i\eta}\right)
\label{atomic-sigma2}
\end{equation}

On the other hand, the retarded self-energy of this model can be calculated up to $\lambda^2$ from the appropriate diagrams using perturbation theory \cite{ISA}. In addition to a constant Hartree contribution this self-energy has the form:
\begin{equation}
\Sigma_{00}^{(2)r}(\omega)=\lambda^2\left(\int_{\mu}^{\infty}d\epsilon\frac{\rho^{(0)}(\epsilon)}{\omega-\epsilon-\omega_0+i\eta}+\int_{-\infty}^{\mu}d\epsilon\frac{\rho^{(0)}(\epsilon)}{\omega-\epsilon+\omega_0+i\eta}\right)
\label{selfret2}
\end{equation}
where $\rho^{(0)}(\omega)=\Gamma/\left[(\omega-\epsilon_{eff})^2+\Gamma^2\right]/\pi$ is the level density of states of the one-electron unperturbed problem, $\epsilon_{eff}$ being an effective level position which can be used for achieving charge consistency between the one-electron and the interacting cases (see \cite{ISA} for details). 
In the limit $\Gamma\rightarrow0$ the above expression tends to
\begin{equation}
\Sigma_{00}^{(2)}(\omega)\rightarrow\lambda^2\left(\frac{1-<\hat{n}>_0}{\omega-\epsilon_{eff}-\omega_0}+\frac{<\hat{n}>_0}{\omega-\epsilon_{eff}+\omega_0}\right)\equiv F(\omega)
\label{selfret2-atomic}
\end{equation}

The interpolative self-energy is then calculated by means of the following \textit{ansatz}:
\begin{equation}
\Sigma_{00}(\omega)=\Sigma^{(at)}\left\{F^{-1}\left[\Sigma_{00}^{(2)}(\omega)\right]\right\}
\label{interpol-equilibrium}
\end{equation}
where $F^{-1}$ is the inverse function defined by Eq. (\ref{selfret2-atomic}). This \textit{ansatz} recovers both the atomic limit $\Gamma/\lambda\rightarrow0$ and the opposite limit where perturbation theory is valid $\lambda/\Gamma\rightarrow0$ and is in excellent agreement with NRG calculations and exact finite system diagonalizations in parameter space \cite{ISA}.

This \textit{ansatz} can be generalized for a nonequilibrium stationary situation like the one addressed in the present work (previous theoretical approaches have been restricted so far to the case of electron-electron interactions \cite{dotne,Aligia}).
In the perturbative limit the self-energies can be calculated up to order $\lambda^2$ using the Keldysh formalism. The second order expressions are \cite{Caroli}:
\begin{eqnarray}
\Sigma_{00}^{(2)+-}(\omega)&=&-i\lambda^2\int\frac{d\nu}{2\pi}G_{00}^{(0)+-}(\omega-\nu)D^{(0)+-}(\nu) \nonumber \\
\Sigma_{00}^{(2)-+}(\omega)&=&-i\lambda^2\int\frac{d\nu}{2\pi}G_{00}^{(0)-+}(\omega-\nu)D^{(0)-+}(\nu)  \nonumber \\
\Sigma_{00}^{(2)r}(\omega)&=&i\lambda^2\int\frac{d\nu}{2\pi}[G_{00}^{(0)r}(\omega-\nu)D^{(0)+-}(\nu)+G_{00}^{(0)-+}(\omega-\nu)D^{(0)r}(\nu)] 
\label{self2-Keldysh}
\end{eqnarray} 
where $D^{(0)}(\omega)$ is the unperturbed phonon propagator and $G_{00}^{(0)}(\omega)$ are the electronic propagators of the quantum level for the nonequilibrium effective one electron problem.

From Eqs. (\ref{self2-Keldysh}) it is straightforward to verify that in the limit $\Gamma\rightarrow0$, $\Sigma_{00}^{(0)r}(\omega)$ tends to an expression formally identical to that of Eq. (\ref{selfret2-atomic}) in the equilibrium situation. Therefore the \textit{ansatz} of Eq. (\ref{interpol-equilibrium}) will recover automatically i) the atomic limit and ii) the results of nonequilibrium perturbation theory in the limit $\lambda/\Gamma\rightarrow0$.   
There still remains the problem of finding an analogous interpolative \textit{ansatz} for the Keldysh self-energies $\Sigma^{+-}$ and $\Sigma^{-+}$ \cite{dotne,Ng,Fazio,Aligia}. 
This is not as straightforward as in the retarded case because these self-energies are not well defined in the atomic limit. An appropriate ansatz can however be obtained by requiring that $\Sigma^{+-}$ and $\Sigma^{-+}$ satisfy the Keldysh relation \cite{dotne, Fazio}:
\begin{equation}
\Sigma_{00}^{+-}(\omega)-\Sigma_{00}^{-+}(\omega)=\Sigma_{00}^{r}(\omega)-\Sigma_{00}^{a}(\omega)=2iIm\Sigma_{00}^{r}(\omega)
\label{sigma-Keldysh}
\end{equation}
and that the results of second order perturbation theory will be recovered in the limit $\lambda/\Gamma\rightarrow0$. This conditions are fulfilled by the following \textit{ansatz}:
\begin{equation}
\Sigma_{00}^{+-}(\omega)=\frac{Im\Sigma_{00}^{r}(\omega)}{Im\Sigma_{00}^{(2)r}(\omega)}\Sigma_{00}^{(2)+-}(\omega)
\label{sigmapm}
\end{equation}
with an analogous expression for $\Sigma_{00}^{-+}(\omega)$. In addition to the above requirements this expression recovers the equilibrium limit:
\begin{equation}
\Sigma_{00}^{+-}(\omega)=2iImG_{00}^{r}(\omega)f(\omega)
\label{sigmapm-equil}
\end{equation}
where $f(\omega)$ is the equilibrium Fermi distribution function of the electrodes.
An analogous \textit{ansatz} to the one of Eq. (\ref{sigmapm}) was used in \cite{dotne,Aligia} for the case of electron-electron interactions.  

Finally we will comment on the self-consistency procedure. We impose consistency in 
the dot charge between the one-electron and the interacting problem. This is achieved by introducing an effective dot level position in the one-electron Hamiltonian. 
As we mentioned at the end of section III, current conservation is not necessarily fulfilled for an approximate solution. In particular this is the case for the ISA.
The self-consistent procedure can be nevertheless generalized by requiring both charge and current consistency between the one-electron and interacting cases \cite{dotne}. A natural choice is to introduce effective one-electron couplings of the dot level with the electrodes $\Gamma_{L,eff},\Gamma_{R,eff}$ which are fixed from the requirement of current consistency. 
When imposing current consistency, the current is calculated by means of Eqs.(\ref{current2}); otherwise Eq.(\ref{current-Wingreen}) is used.
In this work and for the range of parameters considered, the requirement of current consistency does not alter in a significant way the results for current and conductance but the agreement with the EOM results somewhat improves when imposing it.

\section{Results and discussion}

All the calculations in this work have been performed in the limit of zero temperature 
so only phonon emission is possible.
Currents are plotted in units of $e/h$ and conductances are plotted in units of $e^2/h$ .

\begin{figure}
\includegraphics[width=8cm]{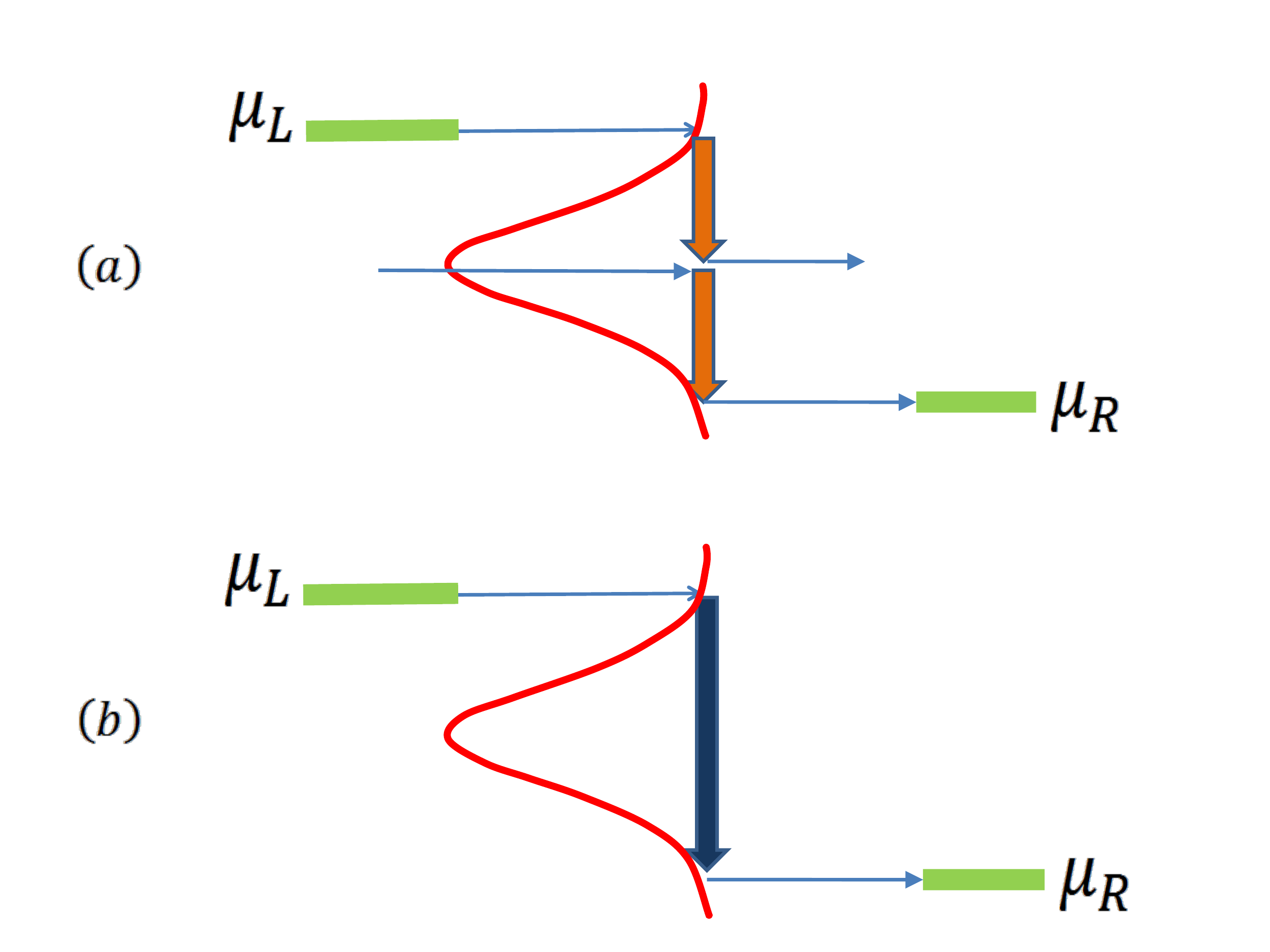}
\caption{(Color online) Sketch of the inelastic processes we analyze in this work. In (a) an 
electron from the left electrode tunnels to level, where it emits $n$ phonons, and passes to
the right electrode.  The onset for emission on $n$ phonons is illustrated in (b), 
where one electron at the left chemical potential tunnels to the level, emits $n$ phonons and 
continues at the right chemical potential. 
Blue thin arrows represent tunneling events and thick arrows phonon emission.
}
\label{fig1}
\end{figure}

It will be useful for the discussion of our results to have a scheme of the inelastic processes 
we describe in this work. Fig.1a sketches a process in which an electron from the left electrode
tunnels to the localized level where it emits $n$ phonons.
Energy conservation requires $\mu_L-\tilde\epsilon_{0}=n\omega_{0}$
(or $\tilde\epsilon_{0}-\mu_R=n\omega_{0}$).
The threshold for emission of $n$ phonons is depicted in Fig.1b. 
It occurs when an electron from the
left electrode jumps into the right electrode through the level and therefore requires
$V=\mu_L-\mu_R=n\omega_{0}$. Both processes show up in the conductance of
the system with characteristic signatures that we analyze in this section.

We start this section by discussing a situation in which the bias potential $V$ is applied symmetrically between the electrodes so that
 $\mu_R=-\mu_L=-eV/2$. The Fermi energy of the leads in equilibrium is taken as our zero of energy.
The symmetry of the problem makes the $I-V$ curves and the conductance to be identical for negative 
and positive values of $\tilde\epsilon_0$. Also $I(-V)=-I(V)$. Then we show results only for 
positive values of  $\tilde\epsilon_0$ and $V$. 

\begin{figure}
\includegraphics[width=14cm]{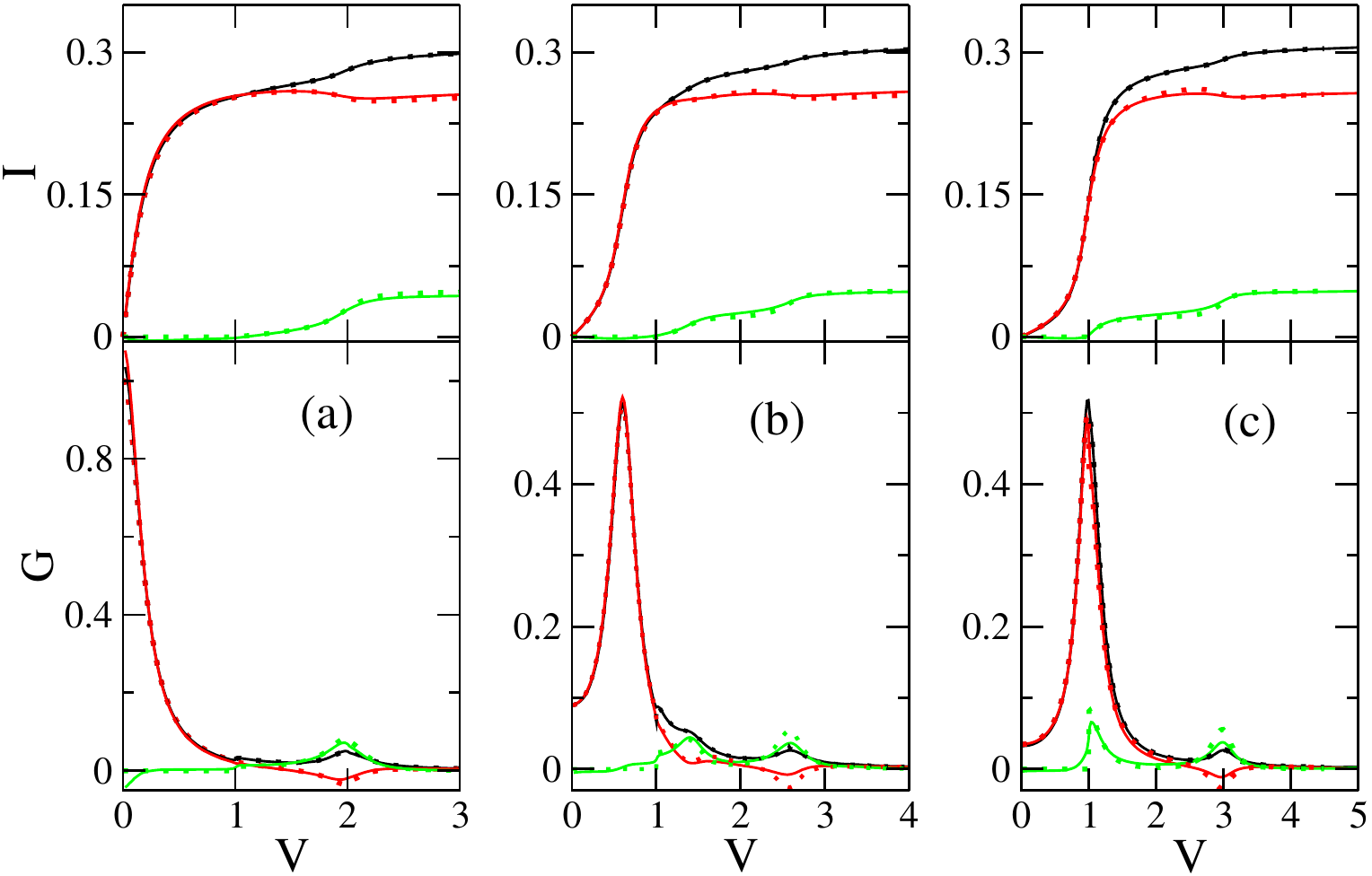}
\caption{ (Color online)
Currents (upper panels) and conductances (lower panels) as a function of $V$
(in units of $\omega_0$) for a symmetrically applied bias voltage and  
$\lambda/\omega_0=0.3$ . Red lines: elastic components,
green lines: inelastic components, black lines: total values. 
Continuous lines: EOM, dotted lines: ISA.
(a) $\tilde\epsilon_0=0$, (b) $\tilde\epsilon_0=0.3 \omega_0$ 
and (c): $\tilde\epsilon_0=0.5 \omega_0$
}
\label{fig2}
\end{figure}

\begin{figure}
\includegraphics[width=14cm]{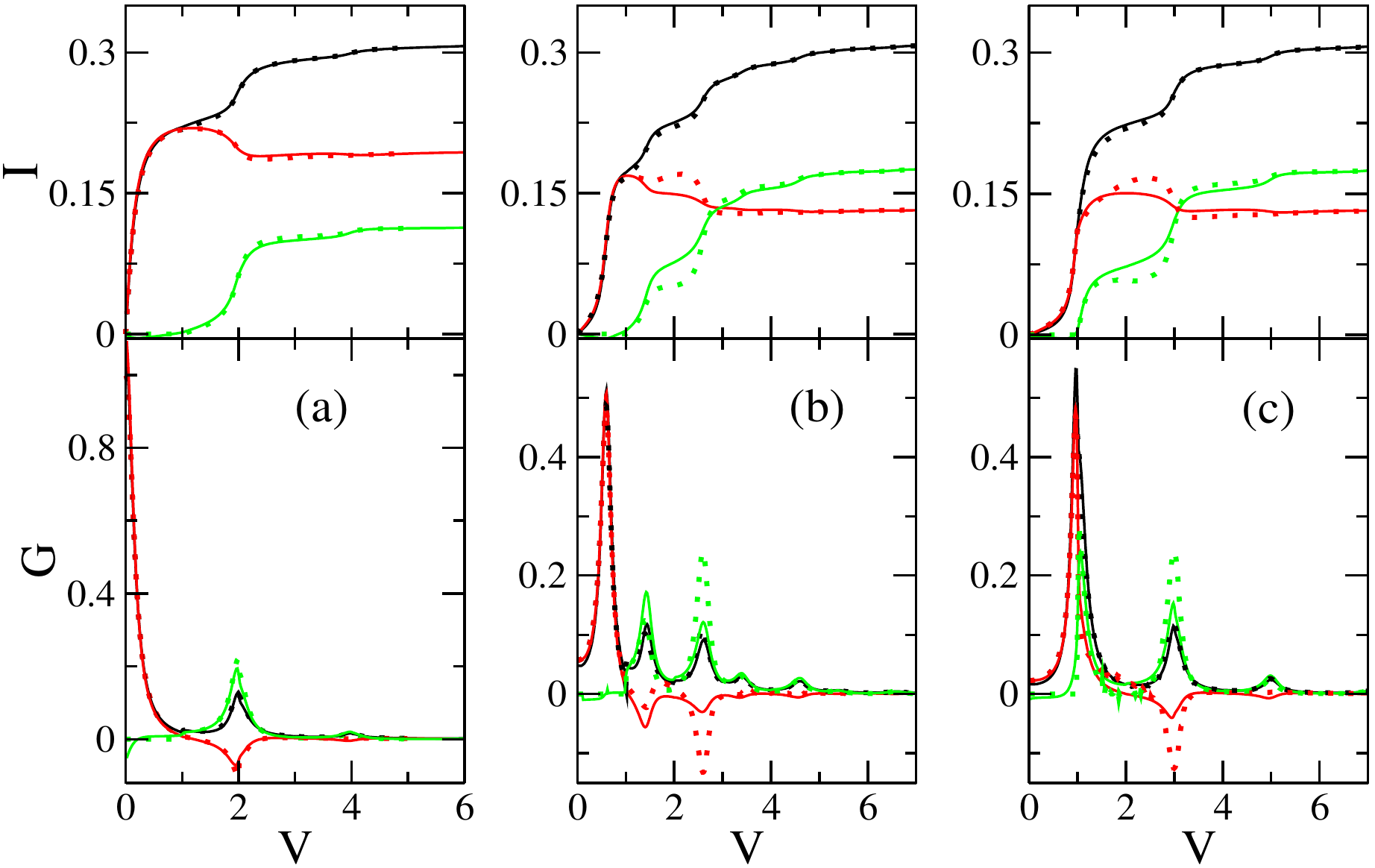}
\caption{ (Color online)
Currents (upper panels) and conductances (lower panels) as a function of $V$
(in units of $\omega_0$) for a symmetrically applied bias voltage and 
$\lambda/\omega_0=0.7$ . Red lines: elastic components,
green lines: inelastic components, black lines: total values. 
Continuous lines: EOM, dotted lines: ISA.
(a) $\tilde\epsilon_0=0$, (b) $\tilde\epsilon_0=0.3 \omega_0$ 
and (c): $\tilde\epsilon_0=0.5 \omega_0$
}
\label{fig3}
\end{figure}

 Fig.2 shows the current (upper panels) and the conductance (lower panels), 
as a function of $V$ for $\lambda=0.3\omega_0$,
$\Gamma=0.1\omega_{0}$
 and for three values of the gate potential corresponding to $\tilde\epsilon_{0}=0.0, 0.3$ and$ 0.5\omega_{0}$. 
The current and the conductance are separated into their elastic and inelastic
contributions, showing clearly that, for this small value of $\lambda/\omega_0$, 
the current is predominantly elastic.
 The inelastic current has a threshold at the onset for inelastic processes, 
$V=\omega_{0}$ for phonon emission which shows up as a step in the conductance. 
Even though this step is tiny on the scale of this figure in cases (a) and (b) because of the
small value of $\lambda/\omega_0$ used here, 
it is a feature that we will discuss extensively
in the context of fig.4.
The conductance also shows different lorentzian-like peaks at
$V=2|\tilde\epsilon_{0} \pm n\omega_{0}|$, with $n$ a positive integer.
 These peaks are the signature of the inelastic processes described in Fig.1a. 
For the case $\tilde\epsilon_{0}=0$ of Fig.2a
they appear at $V=0, 2\omega_0, \; 4\omega_0 \;... $ while for $\tilde\epsilon_{0} > 0$
each  peak is split into two which, according to the energy conservation requirements 
stated above, appear at 
$V=2|\tilde\epsilon_{0} \pm n\omega_{0}|$.
The peaks of the conductance 
correspond to the steps in the $I-V$ curves, their width being proportional to $\Gamma$.     

Fig.3 is as Fig.2 but we have increased the value of the electron-phonon 
interaction to $\lambda=0.7 \omega_0$ (while keeping the same values of 
$\tilde\epsilon_0$). For this value of $\lambda$ we are far from the perturbative regime and the steps in the $I-V$ curve for $n=2$ are clearly visible. Notice how 
the contribution of the inelastic processes to the total current and the conductance 
increases quickly with the applied bias, overcoming the contribution of the elastic processes, 
as we move away from the electron-hole symmetric case 
$\tilde\epsilon_{0}=0$. 
 This behavior,   
in which the current versus voltage curves 
tend to adopt a staircase form with steps located at 
$V=2|\tilde\epsilon_{0} \pm n\omega_{0}|$, is enhanced as 
$\lambda/\omega_{0}$ gets larger than 1.
The height of the steps in the current gives the probability of emitting $n$ phonons 
and follows very approximately the Poisson distribution, $e^{-g}\frac{g^n}{n!}$, with
 $g=(\frac{\lambda}{\omega_0})^2$. This behavior is qualitatively similar to what was obtained in
Ref.\cite{Mitra} using a semiclassical master equations approach. The staircase behavior of conductance with applied bias due to phonon emission has been experimentaly found in Ref\cite{Zhit}. 
The main peak of the conductance, obtained at low voltages, is almost identical 
to the main resonance appearing in the equilibrium density of states, showing the polaronic
reduction of the level width \cite{EOM}. 
However, the phonon side-bands show specific features associated
to inelastic transport, which we will analyze next.
The total conductance shows steps at $V=n\omega_0$. 
We should mention that not only the inelastic  component exhibits this feature but
the elastic component as well 
because of the change in the
retarded self-energy due to the appearance of new inelastic processes.

In Figs. 2 and 3 we compare the results from both theoretical approaches, EOM and ISA. 
The remarkable agreement found gives confidence in the interpolative scheme and also 
in the EOM method to the order $O(V_{k,\nu}^2)$ for values of $\lambda/\omega_{0}$ up to 1.
At this point we should comment that the EOM method up to the order $O(V_{k,\nu}^2)$
starts to show numerical instabilities for higher values of $\lambda$ associated with the 
increasing number of phonons that have to be included in the solution of 
Eqs. (\ref{eomGa}) and (\ref{eomF}) and with the corresponding
 logarithmic singularities in $\Gamma^{(e),(h)}$ (Eq.(\ref{Gammas})). This problem was already 
found in equilibrium and it is cured by the renormalization of these singularities
that appears when the method is carried to the order 
$O(V_{k,\nu}^4)$. 
However, the extension of the procedure to situations out of equilibrium is not straightforward 
and will be deferred to further work.

As mentioned in the Introduction, 
the issue of whether the steps in the total conductance at $V=\omega_{0}$ 
are upwards or downwards has raised a great interest both theoretically and experimentally. Both our formalisms recover the results already obtained in
the weak coupling regime and in the following we concentrate in the regime of strong coupling, 
$\lambda,\omega_{0} > \Gamma$, where we find jumps of the conductance 
at $V=n\omega_0$ for any $n$. 

\begin{figure}
\includegraphics[width=12cm]{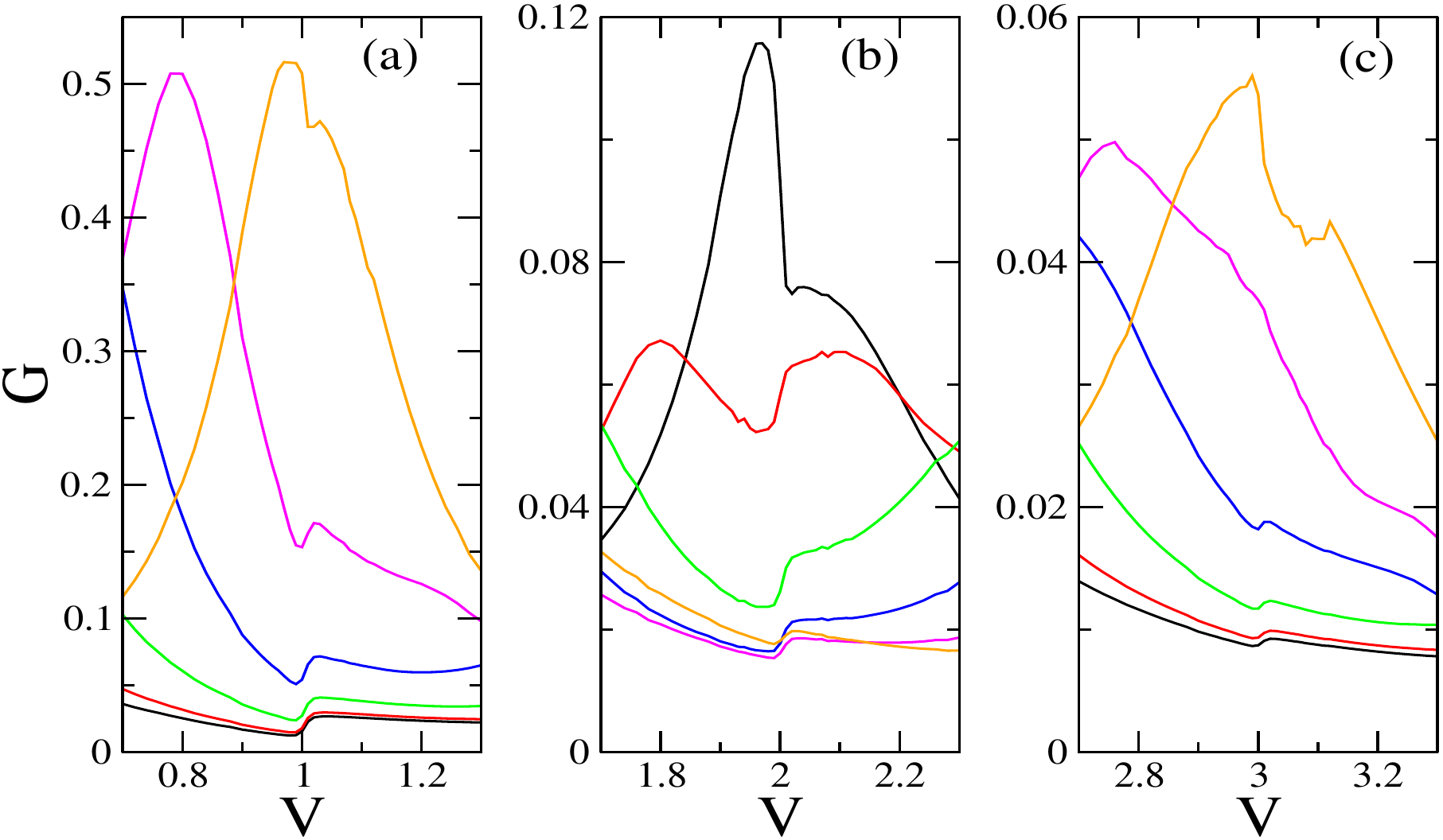}
\caption{ (Color online)
The conductance as a function of $V$ (in units of $\omega_0$) for 
a symmetrically applied voltage and $\lambda/\omega_0=0.5$.
Panels (a), (b) and (c) show the regions near $V=\omega_0$, $V=2 \omega_0$
and $V=3 \omega_0$ respectively.
The results of the EOM method are shown 
for: $\tilde\epsilon_0=0$ (black lines), $\tilde\epsilon_0=0.1\omega_0$ (red lines),
$\tilde\epsilon_0=0.2\omega_0$ (green lines), $\tilde\epsilon_0=0.3\omega_0$ (blue lines)
$\tilde\epsilon_0=0.4\omega_0$ (magenta lines), $\tilde\epsilon_0=0.5\omega_0$ (orange lines).
}
\label{fig4}
\end{figure}

 Fig.4 shows the conductance as a function of the applied bias voltage in the regions near: 
(a) $\omega_{0}$,
(b) $2\omega_{0}$ and (c) $3\omega_{0}$, for $\Gamma=0.1\omega_{0}$, $\lambda=0.5 \omega_{0}$ 
and several values of the gate voltage corresponding to 
$\tilde\epsilon_{0}=0, 0.1, 0.2, 0.3, 0.4$ and $0.5 \omega_{0}$ .
For the sake of clarity, only the results of 
the calculations using the EOM method are shown.
Note in Figs.4(a) and (c) that the step in the conductance is always upwards except 
for $\tilde\epsilon_{0}=0.5 \omega_{0}$, where it is downwards and the conductance is at a relative
maximum.
The same happens in Fig.4(b), with the conductance jumping downwards only 
for $\tilde\epsilon_{0}=0$, for which value the conductance has a relative maximum 
at $V=2\omega_{0}$ . 
These results can be understood in terms of the interference between the step-like processes at
$V=n\omega_0$ and the lorentzian-like peaks at $V=2|\tilde\epsilon_{0} \pm n'\omega_{0}|$.
When both conditions do not coincide, the inelastic conductance increases at $V=n\omega_0$  and
dominates the elastic decrease
which is very small there. Consequently, the conductance
step is upwards. However, if $2|\tilde\epsilon_{0} \pm n'\omega_{0}|=n\omega_0$ (within
an accuracy of $\pm \Gamma$), we always find a downward  decrease of the total conductance steps.
The origin of this behavior is different for $n=1$ than for the rest of the cases. 
The value $V=\omega_0$ is the absolute onset for inelastic processes and, consequently, the 
inelastic conductance increases there. This increase is compensated by a stronger decrease of the elastic
conductance in a way similar to the one analyzed theoretically in the perturbative regime
$\Gamma>>\lambda, \omega_0$ \cite{Vega,Danish,Egger}. 
However, for $n>1$ we find the inelastic conductance decreasing at
$V=n\omega_0$ while the elastic one increases there. 
The appearance of a new inelastic channel emitting $n$ phonons makes 
the intensity of the previously existing ones to decrease abruptly.
Thus we attribute the different behaviors
of the elastic/inelastic components of the conductance to interferences between the 
inelastic processes of Figs. 1a and b, which can occur for $n=2,3...$. The total
conductance always shows a downward step whenever the value $V=n\omega_0$ is at a relative maximum.
In any other case, the conductance jumps up at 
$V=n\omega_0$. This seems to be a very general behavior, valid in both the strong and weak
coupling regimes  in $\lambda/\Gamma$, at least in cases of symmetric coupling between
the localized level and the electrodes. It is seen for any value of $\lambda$ 
not only for $n=1$, as the perturbation theory predicts, but for any value of $n$.

\begin{figure}
\includegraphics[width=8cm]{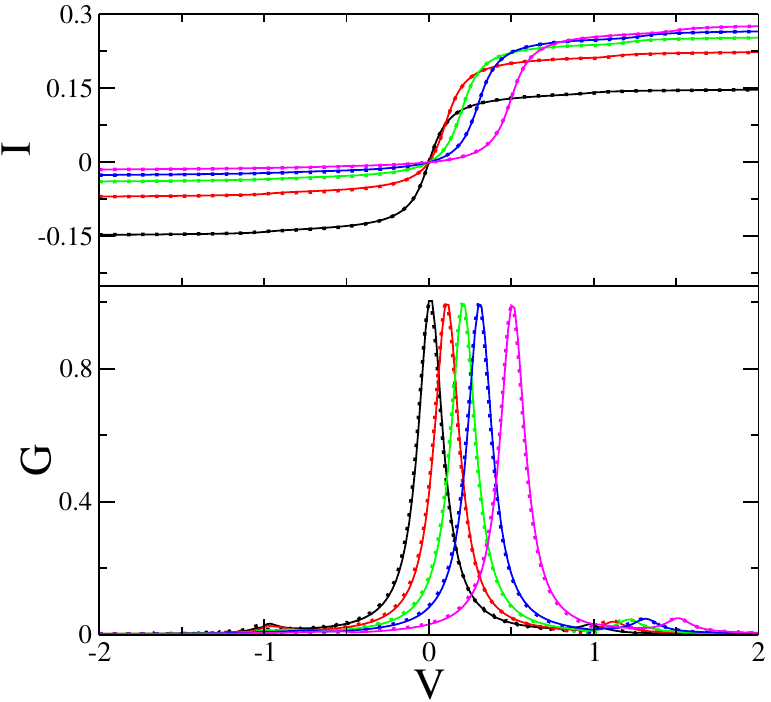}
\caption{ (Color online)
Currents (upper panel) and conductances (lower panel) as a function of $V=\mu_L$
(in units of $\omega_0$) for $\lambda/\omega_0=0.3$ and:
$\tilde\epsilon_0=0$ (black lines), $\tilde\epsilon_0=0.1 \omega_0$ (red lines),
$\tilde\epsilon_0=0.2 \omega_0$ (green lines), $\tilde\epsilon_0=0.3 \omega_0$ (blue lines)
and $\tilde\epsilon_0=0.5 \omega_0$ (magenta lines).
Continuous lines: EOM, dotted lines: ISA.
}
\label{fig5}
\end{figure}

\begin{figure}
\includegraphics[width=8cm]{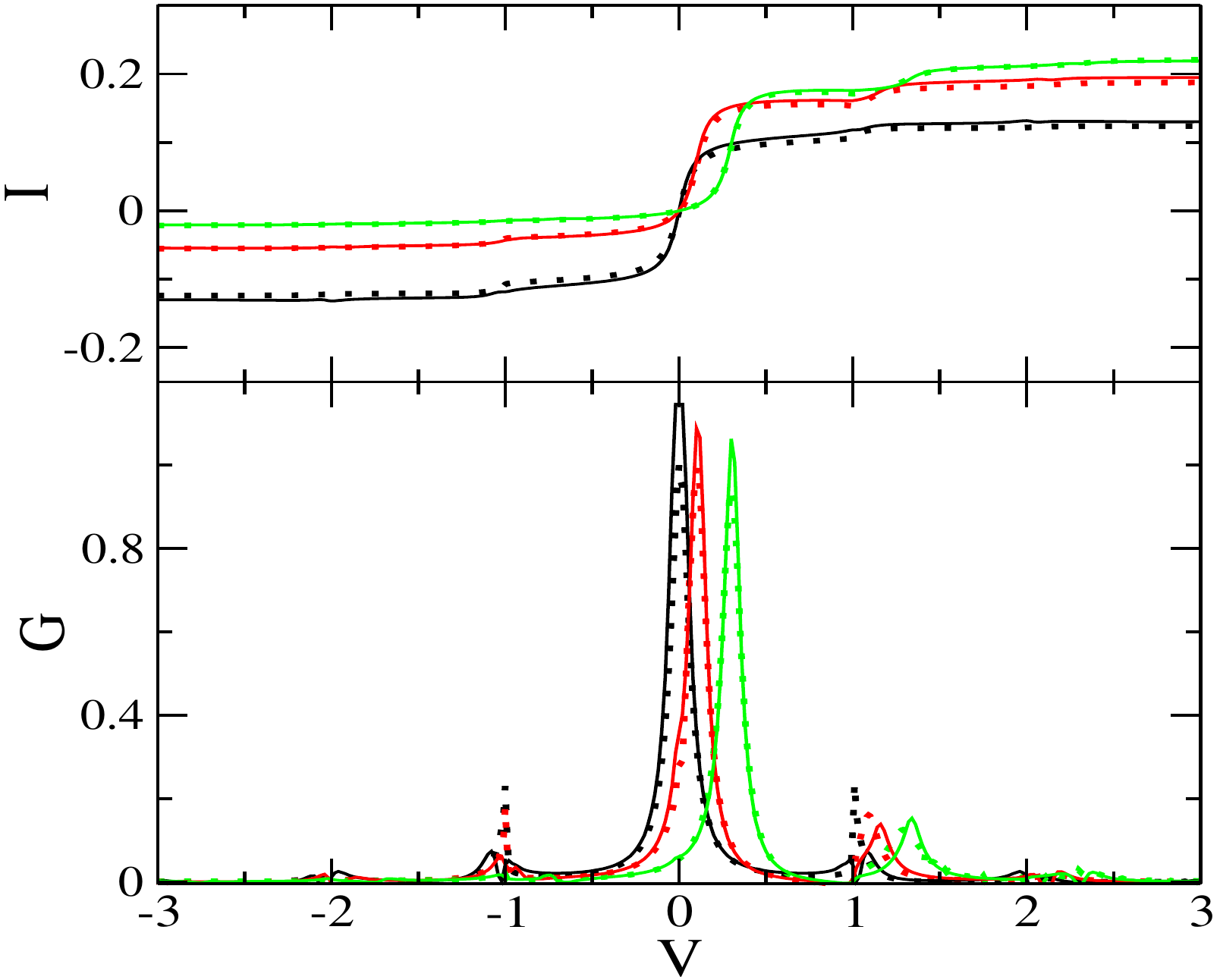}
\caption{ (Color online)
Currents (upper panel) and conductances (lower panel) as a function of $V=\mu_L$
(in units of $\omega_0$) for $\lambda/\omega_0=0.7$ and:
$\tilde\epsilon_0=0$ (black lines), $\tilde\epsilon_0=0.1 \omega_0$ (red lines),
$\tilde\epsilon_0=0.3 \omega_0$ (green lines)
Continuous lines: EOM, dotted lines: ISA.
}
\label{fig6}
\end{figure}

We have already pointed out the good agreement obtained by our two theoretical approaches 
in the case of a symmetrically applied bias. That this agreement is not fortuitous 
is proved by comparing the results in a different situation, 
in which the bias is applied asymmetrically, with 
$\mu_R=0$ and $\mu_L=V$. This is done in Figs.5 and 6, where we show the current and the conductance  for 
$\lambda=0.3\omega_{0}$ and $\lambda=0.7\omega_{0}$ respectively, for several values of  
$\tilde\epsilon_{0}$.
For simplicity, we have chosen $\Gamma_L=\Gamma_R=\Gamma/2$
with $\Gamma=0.1 \omega_{0}$.  
Only positive values of $\tilde\epsilon_{0}$ are shown because 
$I(-\tilde\epsilon_{0},V)=-I(\tilde\epsilon_{0},-V)$ and
$G(-\tilde\epsilon_{0},V)=G(\tilde\epsilon_{0},-V)$. 
At variance from Figs.2 and 3, the maximum of the conductance is very close to 1. 
The larger deviations from perfect conductance are obtained 
in Fig.6, for large  $\tilde\epsilon_{0}$ which means that we are far from equilibrium. The fact that the EOM results are higher than the interpolative results
at the maximum is the consequence of the numerical inaccuracies commented above.
As in Figs.2 and 3, the current increases in a step-like way. Correspondingly,
the conductance presents lorentzian-like phonon side-bands 
 associated with the inelastic process occurring at
$V=\tilde\epsilon_{0} \pm n\omega_{0}$ and jumps at $V=m\omega_{0}$, with a strong change in line shape under conditions 
when they can both occur and interfere. 
Therefore, this is a robust behavior obtained by both theoretical approaches under different 
values of the parameters defining the problem. 
 The asymmetry of the conductance for positive and negative values of $V$ is a
consequence of the very asymmetric behavior of the level occupancy $<n_{0}(V)>$ when one
of the electrodes do not change its chemical potential.  This
can be qualitatively understood from the atomic Green function, Eq. (\ref{atomic-G}), 
where one can readily see that, for
positive values of $\tilde\epsilon_{0}$ and $\omega_{0}>\tilde\epsilon_{0}$, 
phonon emission with $V>0$ ($V<0$) 
should be proportional to $1-<n_0>$ ($<n_0>$). Also, the asymmetry of the conductance follows the shape of the nonequilibrium density of states (not shown) with $V>0$ ($V<0$) mapping out its empty (occupied) portions.

\section{Conclusions}
In this work, we present a combined theoretical approach to analyze the nonequilibrium transport properties of nanoscale systems exhibiting strong electron-phonon interactions and coupled to metallic electrodes. 
We describe the system by the spinless Anderson-Holstein Hamiltonian and 
 use a Keldysh Green function formalism to generalize an equation of motion method and an
interpolative self-energy approximation previously developed in equilibrium. 
These two approaches recover the results obtained formerly in the weak coupling regime 
$\lambda/\Gamma<<1$
and this article concentrates in the strong coupling regime $\lambda, \omega_0>\Gamma$.
Using both techniques, 
we analyze the specific features of inelastic transport appearing in the intensity 
versus voltage 
curves and in the conductance. 
Excellent overall agreement between both approaches is found
in a wide range of parameters. 
We obtain a step-like increase of the current with the applied
voltage at $V=\tilde\epsilon_{0} \pm n\omega_{0}$
with the corresponding phonon sidebands of the conductance, a behavior which gets more
pronounced as $\lambda/\omega_{0}$ increases. 
We also find steps in the conductance at $V=n\omega_0$ for any value of $n$.
These are generally upwards, except when the value $V=n\omega_0$ occurs at a relative maximum
of the conductance in which case it is downwards.
This seems to be a very general behavior, valid in both the strong and weak
coupling regimes  in $\lambda/\Gamma$, at least in cases of symmetric coupling between
the localized level and the electrodes.  

\begin{acknowledgements}      
We thank J.M. Benavides for drawing Fig1. 
Support by the Spanish Ministerio de Ciencia e Innovaci\'on
contracts FIS2008-04209 and MAT2007-60966, and by the Comunidad Aut\'onoma de Madrid, project Nano-objects S2009/MAT-1467, 
is acknowledged.
\end{acknowledgements}


\end{document}